\documentclass[prx,preprint,twoside,showpacs]{revtex4-1}
\usepackage{graphics,amssymb,amsmath,epsf,color,hyperref}
\epsfclipon
\usepackage{epsfig,bm,epsf,graphics}
\usepackage{graphicx}% Include figure files
\usepackage{dcolumn}% Align table columns on decimal point
\usepackage{bm}% bold math
\usepackage{color}
\usepackage{soul}
\usepackage{xcolor}

\begin{document}
\title{Statistical Physics of Epidemic on Network Predictions for SARS-CoV-2 Parameters}

\author{Jungmin Han}%
\author{Evan C Cresswell-Clay}

\author{Vipul Periwal}
\email[Corresponding author: ]{vipulp@mail.nih.gov}
\affiliation{Laboratory of Biological Modeling, National Institute of Diabetes and Digestive and Kidney Diseases, National Institutes of Health, Bethesda, Maryland 20892, USA}

\date{\today}

\begin{abstract}
The SARS-CoV-2 pandemic has necessitated unprecedented mitigation efforts around the world. % but basic parameters of the disease are still being determined. %Reports that age is a strong factor in mortality, that infectivity may sometimes lasts for long periods or that infected individuals may sometimes be asymptomatic are a strong impetus for ascertaining such parameters as accurately as feasible. 
Our aim in this study was to use only reported deaths in the early days before mitigation efforts are initiated to determine basic infection parameters, in order to make credible predictions of hidden variables such as the time dependence of the number of infections. The early deaths are sporadic and discrete so the use of network models of epidemic spread is imperative, with stochasticity in the network itself a crucial random variable. Furthermore, location-specific population age distributions and population densities must be taken into account when attempting to fit these early events with parametrized models. These characteristics combine to render a garden-variety Bayesian model comparison impractical as the networks have to be large enough to avoid finite-size effects. %proportional to $\mathrm{Age}^{5.5}$ (based on New York state data) 
We reformulated this problem as the statistical physics of independent location-specific `balls' attached to every model in a six-dimensional lattice of 56448 parametrized models by elastic springs, with model-specific `spring constants' determined by the stochasticity of network epidemic simulations for that model. The distribution of balls then determines all Bayes posterior expectations. 
Important characteristics of the contagion are determinable: the fraction of infected patients that die ($0.017\pm 0.009$), the expected period an infected person is contagious ($22 \pm 6$ days)  and the expected time between the first infection and the first death ($25 \pm 8$ days) in the US. 
The rate of exponential increase in the number of infected individuals is $0.18\pm 0.03$ per day, corresponding to 65 million infected individuals in one hundred days from a single initial infection. With even imperfect social distancing effectuated two weeks after the first recorded death, this number fell to 166000 infected individuals. We predict that the fraction of compliant socially-distancing individuals matters less than their fraction of social contact reduction for altering the cumulative number of infections and deaths after the effectuation of social distancing. \end{abstract}

%\pacs{05.45.Xt, 05.45.-a, 05.70.Fh}
%05.45.Xt: coupled oscillators, synchronization, nonlinear dynamics
%05.45.-a: dynamical systems, nonlinear dynamics
%05.70.Fh: phase transition in statistical mechanics and thermodynamics
 
 \maketitle

%\tableofcontents

\section{Introduction}
The pandemic caused by the SARS-CoV-2 virus has swept across the globe with remarkable rapidity. The parameters of the infection produced by the virus, such as the infection rate from person-to-person contact, the mortality rate upon infection and the duration of the infectivity period are still controversial \cite{berger2020seir, calvetti2020bayesian, Kucharski2020, zhang2020evolving, lourenco2020fundamental, chow2020global, chang2020modelling, keeling2020predictions, seydi2020unreported, raghavan2020using, hota2020closedloop, singh2020c19tranet, mao2020datadriven, liu2020distributed, susanto2020infect, whiteley2020inference, johansson2020masking, tuite2020mathematical, maltezos2020methodology, plata2020simulating, ganesan2020spatiotemporal, ohsawa2020stay, bastos2020covid19, endo2020estimating, furuse2020clusters, lau2020characterizing}.  Parameters such as the duration of infectivity and  predictions such as the number of undiagnosed infections could be useful for shaping public health responses as the predictive aspects of model simulations are possible guides to pandemic mitigation \cite{raghavan2020using, plata2020simulating, chang2020modelling}. In particular, the possible importance of superspreaders should be understood \cite{reich2020modeling, endo2020estimating, furuse2020clusters, lau2020characterizing}. 

\cite{lourenco2020fundamental} had the insight that the early deaths in this pandemic could be used to find some characteristics of the contagion that are not directly observable such as the number of infected individuals. This number is, of course, crucial for public health measures. The problem is that standard epidemic models with differential equations are unable to determine such hidden variables as explained clearly in \cite{chow2020global}. 

The early deaths are sporadic and discrete events. These characteristics imply that simulating the epidemic must be done in the context of network models with discrete dynamics for infection spread and death. The first problem that one must contend with is that even rough estimates of the high infection transmission rate and a death rate with strong age dependence imply that one must use large networks for simulations, on the order of $10^5$ nodes, because one must avoid finite-size effects in order to accurately fit the early stochastic events. The second problem that arises is that the contact networks are obviously unknown so one must  treat the network itself as a stochastic random variable, multiplying the computational time by the number of distinct networks that must be simulated for every parameter combination considered. The third problem is that there are several characteristics of SARS-CoV-2 infections that must be incorporated in any credible analysis, and the credibility of the analysis requires an unbiased sample of parameter sets. These characteristics are the strong age dependence of mortality of SARS-CoV-2 infections and a possible dependence on population density which should determine network connectivity in an unknown manner. Thus the network nodes have to have location-specific population age distributions incorporated as node characteristics and the network connectivity itself must be a free parameter. 
 
An important point in interpreting epidemics on networks is that the simplistic notion that there is a single rate at which an infection is propagated by contact is indefensible. In particular, for the SARS-CoV-2 virus, there are reports of infection propagation through a variety of mucosal interfaces, including the eyes. Thus, while an infection rate must be included as a parameter in such simulations, there is a range of infection rates that we should consider. Indeed, one cannot make sense of network connectivity without taking into account the modes of contact, for instance if an individual is infected during the course of travel on a public transit system or if an individual is infected while working in the emergency room of a hospital. One expects that network connectivity should be inversely correlated with infectivity in models that fit mortality data equally well but this needs to be demonstrated with data to be credible, not imposed by fiat. The effective network infectivity, which we define as the product of network connectivity and infection rate, is the parameter that needs to be reduced by either social distancing measures such as stay-at-home orders or by lowering the infection rate with mask wearing and hand washing. 

A standard Bayesian analysis with these features is computationally intransigent. We therefore adopted a statistical physics approach to the Bayesian analysis. We imagined a six-dimensional lattice of models with balls attached to each model with springs. Each ball represents a location for which data is available and each parameter set determines a lattice point. The balls are, obviously, all independent but they face competing attractions to each lattice point. The spring constants for each model are determined by the variation we find in stochastic simulations of that specific model. One of the dimensions in the lattice of models corresponds to a median age parameter in the model. Each location ball is attracted to the point in the median age parameter dimension that best matches that location's median age, and we only have to check that the posterior expectation of the median age parameter for that location's ball is  close to the location's actual median age. Thus we can decouple the models and the data simulations without having to simulate each model with the characteristics of each location, making the Bayesian model comparison amenable to computation. Finally, the distribution of location balls over the lattice determines the posterior expectation values of each parameter.

We matched the outcomes of our simulations with data on the two week cumulative death counts after the first death using Bayes' theorem to obtain parameter estimates for the infection dynamics. We used the Bayesian model comparison to determine posterior expectation values for parameters for three distinct datasets. Finally, we simulated the effects of various partially effective social-distancing measures on random networks and parameter sets given by the posterior expectation values of our Bayes model comparison.

\section{Data}
\begin{figure*}
\centering
\includegraphics[width=17cm]{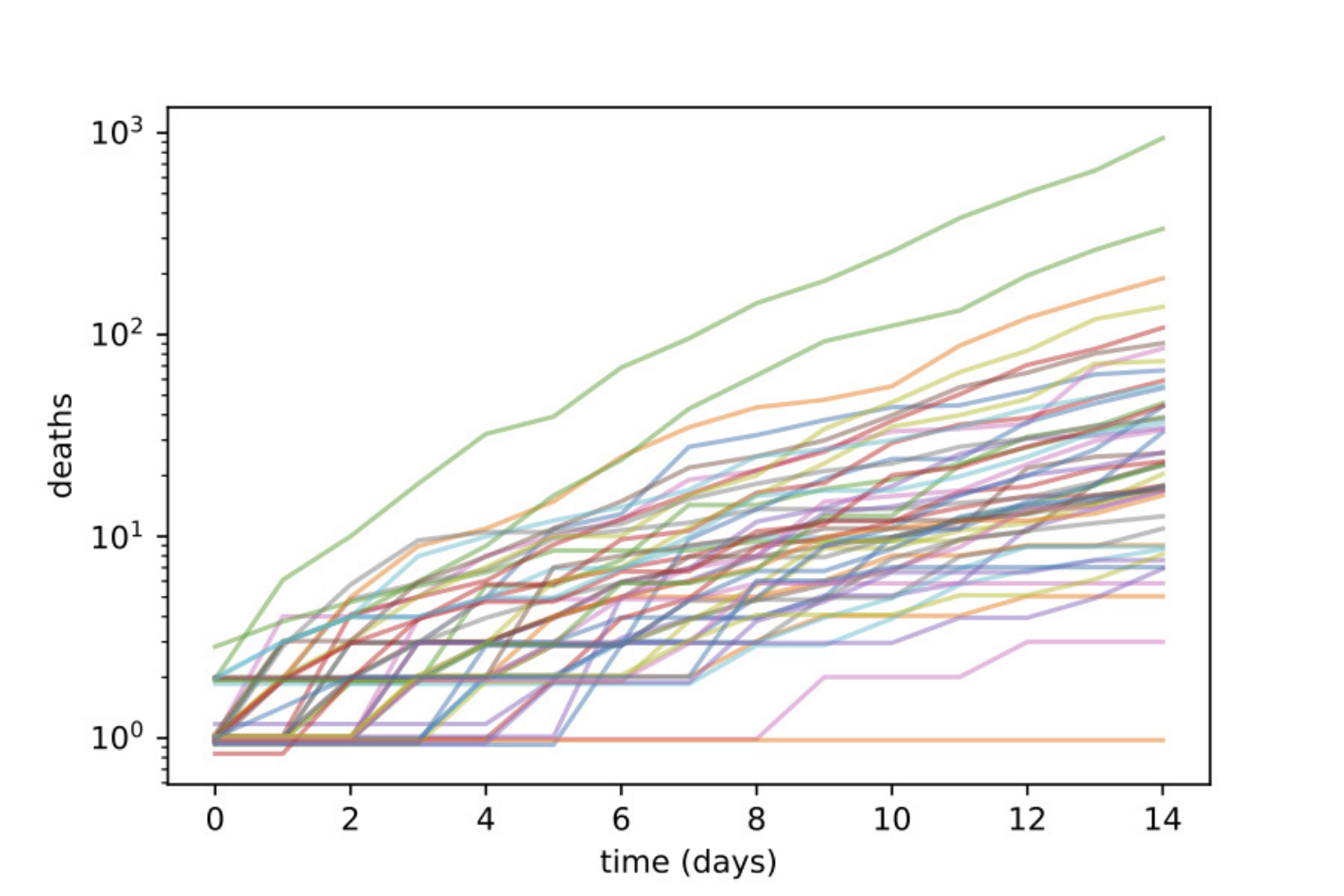}
\caption{ \label{fig:fig1} Cumulative deaths for the first 14 days after the first death in 50 US states and the District of Columbia.
}
\end{figure*}

\begin{figure*}
\centering
\includegraphics[width=17cm]{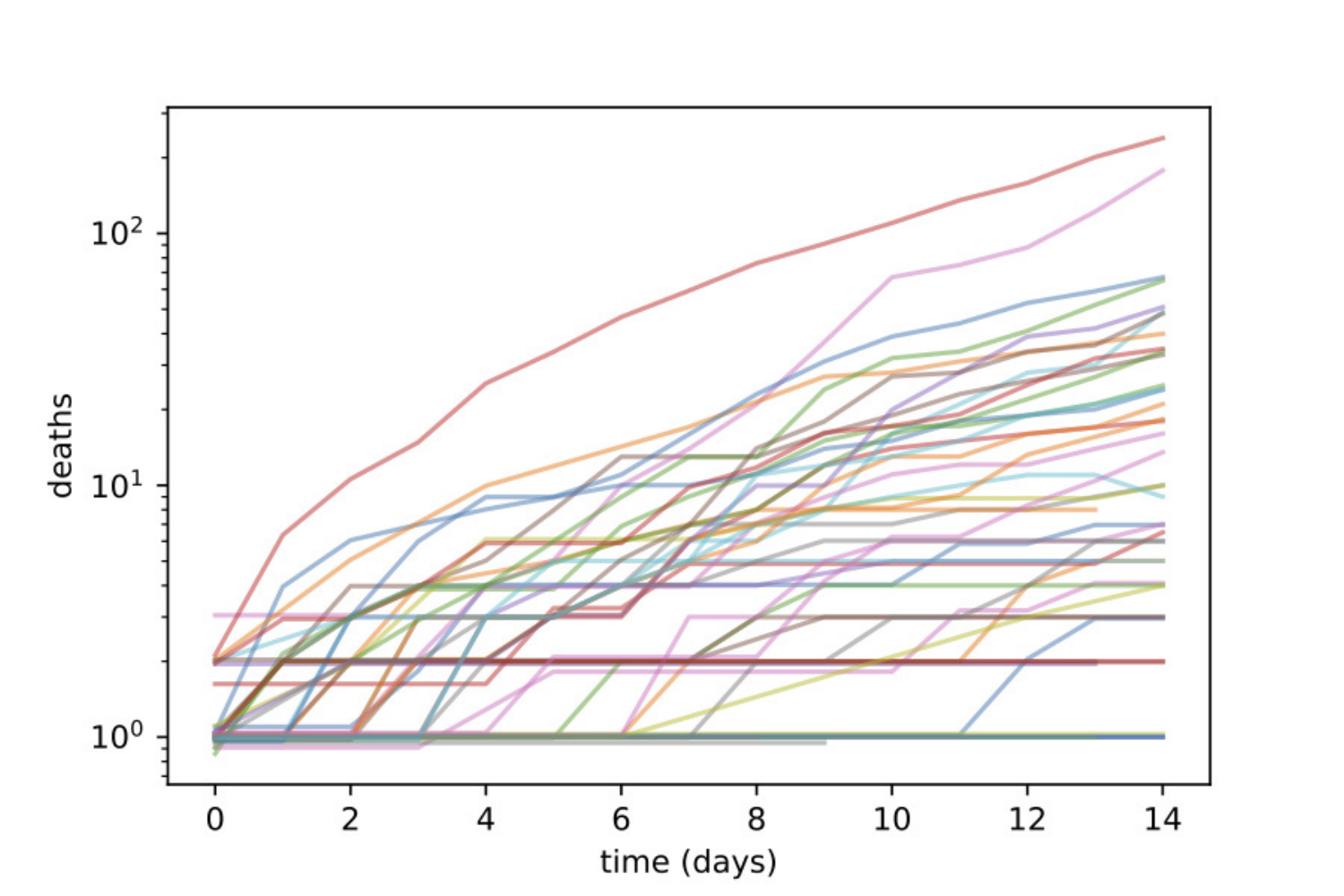}
\caption{ \label{fig:fig2} Cumulative deaths  for the first 14 days after the first death in countries of the world.
}
\end{figure*}

We used data for the SARS-CoV-2 pandemic as compiled by  \cite{datasource} from the original data collected by the New York Times (US state data) (Figure ~\ref{fig:fig1})  \cite{nytdata}, Dong, Du and Gardner at the Johns Hopkins University (Figure~\ref{fig:fig2}) \cite{dong2020interactive}, and the European Centre for Disease Prevention and Control \cite{ecdc}. US state population and median age was obtained from  \cite{USPopMedian} and US state land area from  \cite{USLandArea}. We used World Bank estimates of country-specific population demographics for 178 countries \cite{worldbank} and age-specific mortality statistics for New York state \cite{nystate05132020}. As evident in Figures \ref{fig:fig1} and \ref{fig:fig2}, most of the data do not show any large or abrupt increase in deaths.

\section{Methods}
\label{methods}
\subsection{Networks}
We generated random $G(N,p = 2L/(N-1))$ networks of $N=90000$ or $100000$ nodes with an average of $L$ links per node using the Python package NetworkX \cite{hagberg2008}.   ScalePopDens $\equiv L$ is one of the parameters that we varied. We compared the posterior expectation for this parameter for a location with the actual population density in an attempt to predict the appropriate way to incorporate measurable population densities in epidemic on network models \cite{Anderson1999,amblard2015models}.
\subsection{Epidemic Simulation}
We used the Python Epidemics on Networks package \cite{kiss2017mathematics,Miller2019} to simulate networks with specific parameter sets. We defined nodes to have status Susceptible, Infected, Recovered or Dead. We started each simulation with exactly one infected node, chosen at random. The simulation has two sorts of events:
\begin{enumerate}
\item An Infected node connected to a Susceptible node can change the status of the Susceptible node to Infected with an infection rate, InfRate. This event is network connectivity dependent. Therefore we expect to see a negative or inverse correlation between InfRate and ScalePopDens.
\item An Infected node can transition to Recovered status with a recovery rate, RecRate, or transition to a Dead status with a death rate, DeathRate. Both these rates are entirely node-autonomous. The reciprocal of the RecRate parameter (RecDays in the following) is the number of days an individual is contagious.
\end{enumerate}
We assigned an age to each node according to a probability distribution parametrized by the median age of each data set (country or state). As is well-known, there is a wide disparity in median ages in different countries. The probability distribution approximately models the triangular shape of the population pyramids that is observed in demographic studies. We parametrized it as a function of age $a$ as follows:
\begin{equation}
P(a|\mathrm{MedianAge}) \propto  {1\over{1 + \frac{5}{3} S (a/\mathrm{MaxAge})^{{10}/{3}}}} - {1\over{1 + \frac{5}{3}S }}.
\end{equation}
Here MedianAge is the median age of a specific country, MaxAge $=100$ y is a global maximum age parameter ($P(a=\mathrm{MaxAge}) = 0$), $S\equiv \mathrm{MaxAge}/\mathrm{MedianAge},$ and the coefficients were numerically optimized using least absolute deviation to fit World Bank population demographics for 178 countries \cite{worldbank} (fitting fractions  of the population 14 yo and under, 15 to 65 yo, and 65 and over, and with the fraction up to MedianAge equal to 0.5), and $P(a)$ summed over all ages is normalized to unity. The fits had standard deviations of about 12\%. 

It is computationally impossible to perform model simulations for the exact age distribution for each location. We circumvented this problem, as detailed in the next subsection (Bayes setup), by incorporating a ScaleMedAge parameter in the model, scaled so that ScaleMedAge = 1.0 corresponds to a median age of 40 years. 
The node age is used to make the DeathRate of any node age-specific in the form of an age-dependent weight:
\begin{equation} 
w(a[n]|\mathrm{AgeIncrease}) \propto (a[n]/\mathrm{MaxAge})^\mathrm{AgeIncrease},
\end{equation}
where $a[n]$ is the age of node $n$   and AgeIncrease $ = 5.5$ is an age-dependence exponent. 
$w(a)$ is normalized so that 
\begin{equation} 
\sum_a w(a|\mathrm{AgeIncrease})P(a|\mathrm{MedianAge}=38.7\mathrm{y}) = 1,
\end{equation}
using the median age of New York state's population 
as the value of AgeIncrease given above was approximately determined by fitting to the observed age-specific mortality statistics of New York state \cite{nystate05132020}. However, we included AgeIncrease as a model parameter since the strong age dependence of SARS-CoV-2 mortality is not well understood, with the normalization adjusted appropriately as a function of AgeIncrease. Note that a decrease in the median age with all rates and the age-dependence exponent held constant will lead to a lower number of deaths.
\begin{figure*}
\centering
\includegraphics[width=17cm]{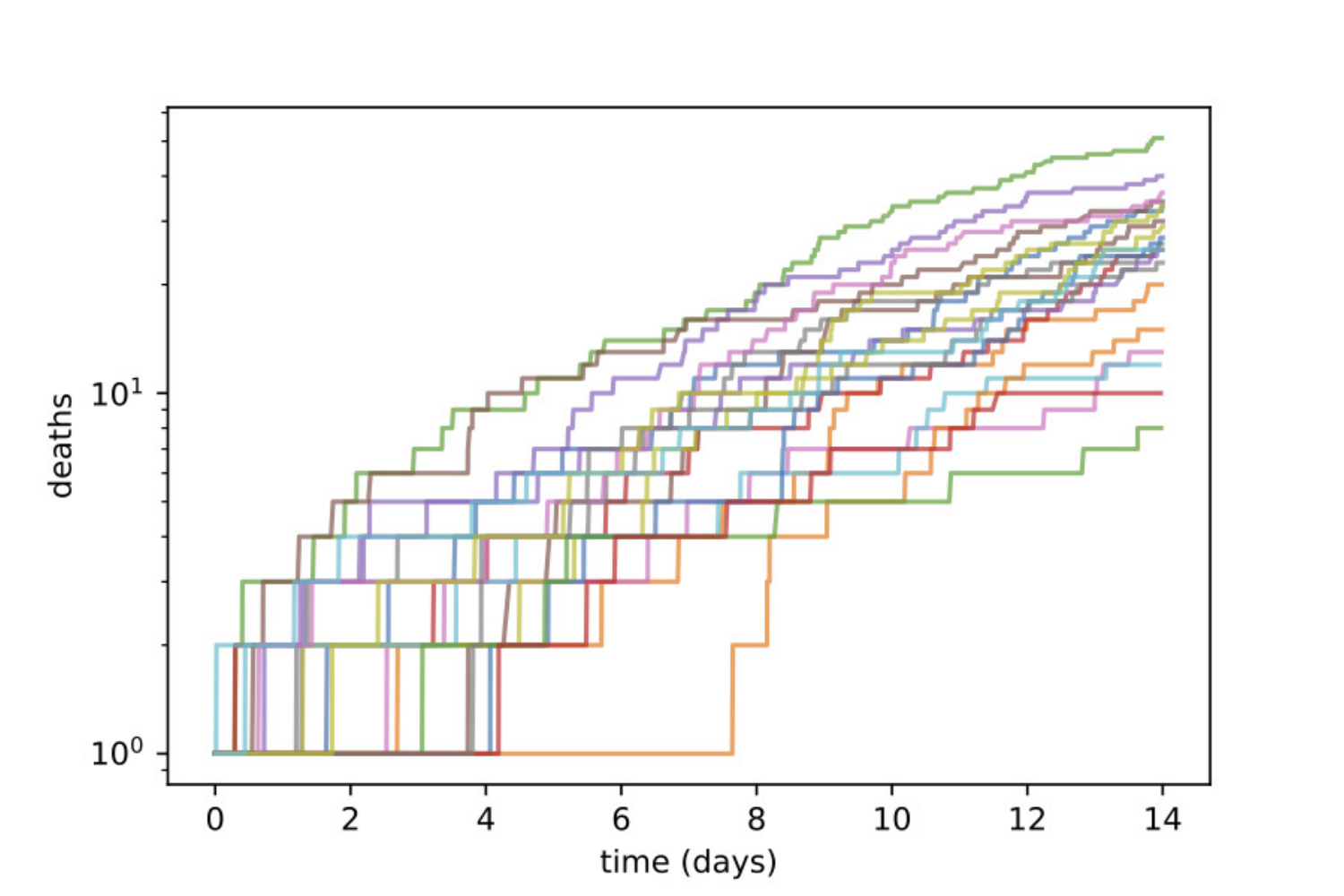}
\caption{ \label{fig:fig3} Example of simulation results showing deaths for the first 15 days including the day of the first death. For fixed $N,p$ values, the simulations have different $G(N,p)$ networks for each simulation, so the network connectivity and the infection spread are both stochastic.
}
\end{figure*}

We use simulations to find the number of Dead nodes as a function of time. The first time at which a death occurs following the initial infection in the network is labeled TimeFirstDeath. Figure~\ref{fig:fig3} shows examples of cumulative number of deaths as a function of time, with all times referred to the time of first death.
\subsection{Statistical physics setup}
\begin{figure*}
\centering
\includegraphics[width=16cm]{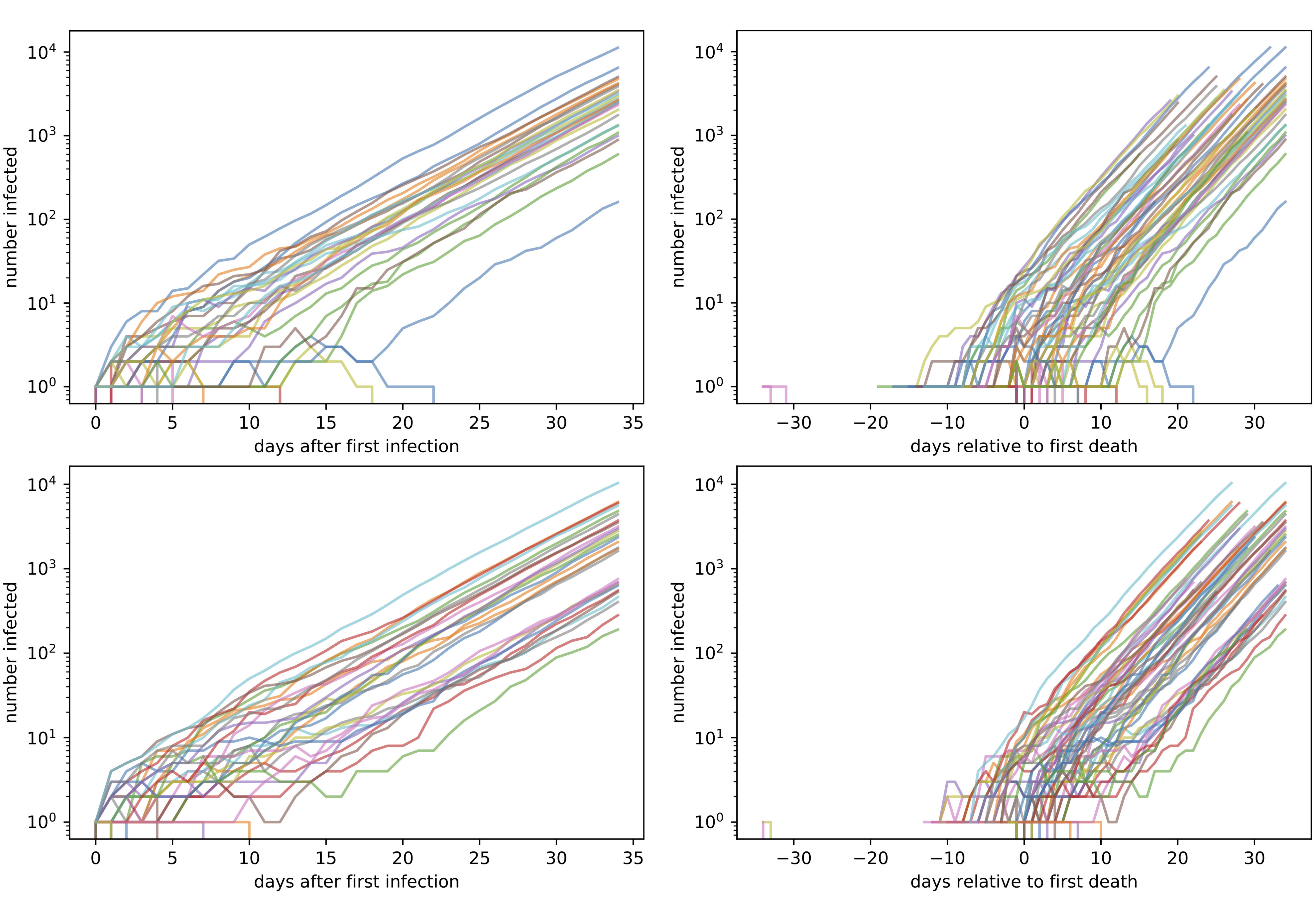}
\caption{ \label{fig:simulations40} An example of 40 simulations for the same parameter set. (Table \ref{tb:post_par}). Upper row: New random graphs for each simulation. Lower row: A fixed random graph and a fixed initial infected node for all simulations. Left panels: The number of infected individuals as a function of time after the first infection. Right panels: The number of infected individuals as a function of time relative to the first death. 
}
\end{figure*}

Given the stochasticity in the simulations for even a fixed network (Figure~\ref{fig:simulations40}) and the unknown structure of the network itself, optimizing parameters is ill-advised. We opted instead to formulate a six-dimensional uniform grid of models (ScalePopDens, InfRate, RecRate, DeathRate, AgeIncrease,  ScaleMedAge) thinking of these as fixed physical centers for spring centers to which data particles are attached. The spring position and minimum energy of each fixed center are complex functions of the six parameters which we determined by 40 repeated model simulations, making sure that our results did not depend on the number of simulations. The initial exponential increase in deaths gives rise to slope and intercept parameters when regressing the logarithm of the cumulative number of deaths as a function of time. By definition, the slope of the logarithm of the cumulative death count is constrained to lie on the positive real line, so we interpreted it as a log-normally distributed variable and modeled its logarithm as normally distributed, calculating the mean and standard deviation of the 40 simulations. The mean value of the logarithm of the slope serves as the spring position center and the standard deviation determines the spring constant and the zero of the potential energy, giving rise to a normalized Gau\ss ian probability distribution. We modeled the intercept as distributed normally  as it has no positivity constraints. 

\begin{figure*}
\centering
\includegraphics[width=17cm]{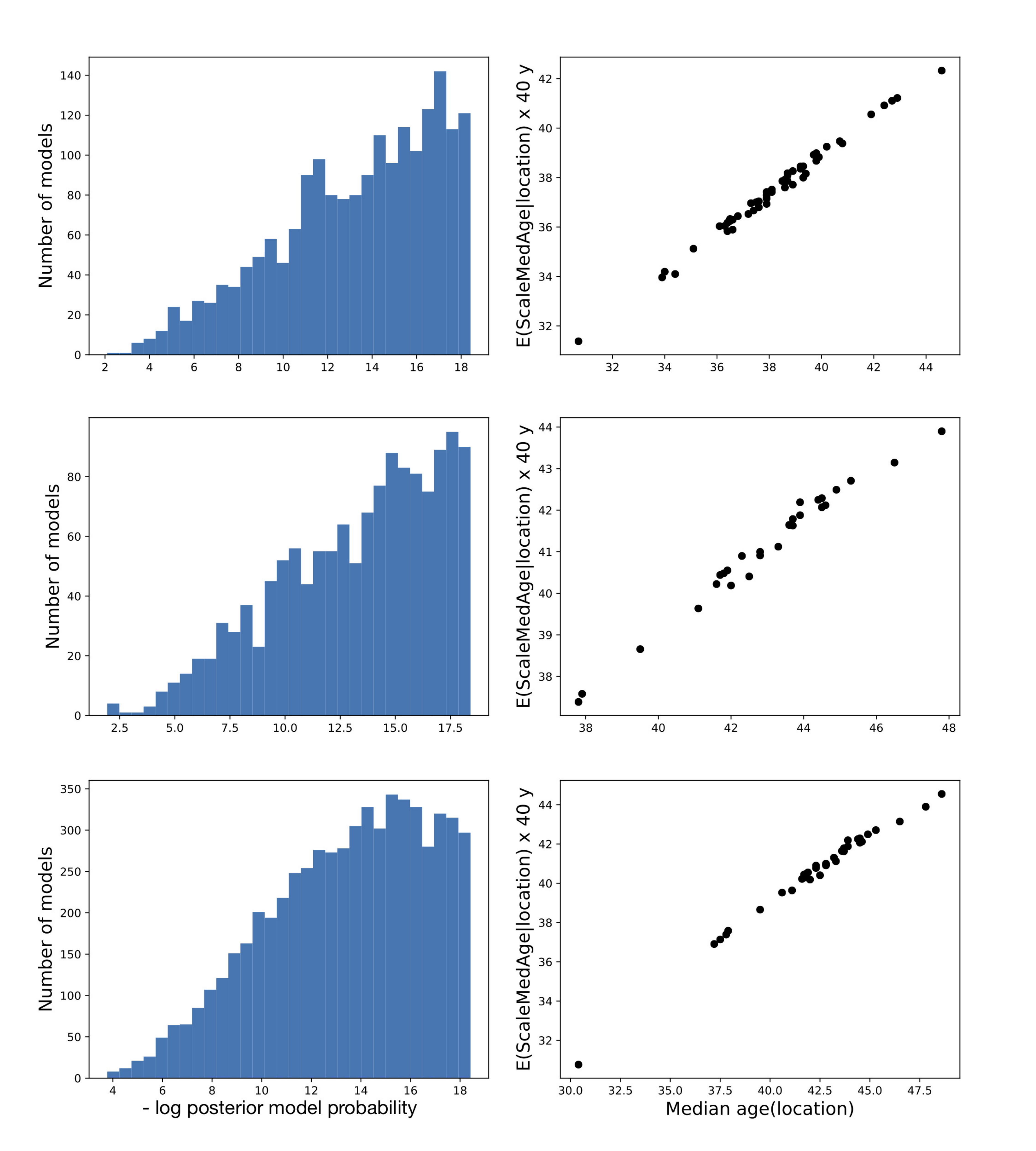}
\caption{ \label{fig:histogram}  Left column: Histogram of model posterior probabilities for models with posterior probabilities $>10^{-8}.$ Right column: Posterior expectation of ScaleMedAge parameter $\times 40$ y vs. Median Age at location.
}
\end{figure*}

With this physical picture in mind, we can take into account the population median age as well by including in the energy an attraction term to the value of the ScaleMedAge parameter of a model center with a spring constant determined by the standard deviation of the worldwide population weighted median age, including the location's median age as part of the location-specific data with a corresponding dimension in the model grid for the model's ScaleMedAge parameter. As we show in the Results (Figure \ref{fig:histogram}), this works well in ensuring that the posterior expectation of a location's ScaleMedAge ($\times 40$ y) parameter is close to  its actual median age.

We implemented Bayes' theorem as usual. The probability of a model, $M,$ given a set of data, $\{D_\alpha\},$ is
\begin{equation}
\mathrm{P}(M|\{D_\alpha\}) = Z^{-1}\prod_\alpha \mathrm{P}(M|D_\alpha) = Z^{-1}\prod_\alpha \exp(-\mathrm{Energy}_M(D_\alpha))
\end{equation}
with our prior probability of any model, $P(M),$ assumed to be a constant independent of $M.$ Here, $\mathrm{Energy}_M(D_\alpha)$ is the potential energy of the data particle $D_\alpha$ (i.e., a state or a country) attached to the springs of model $M$ with terms corresponding to the logarithm of the model slope, the intercept and the ScaleMedAge parameter including constants for normalization for each $M,$ and $Z$ is the normalization partition sum over all models for all $D_\alpha.$ The energy of each data particle in the potential of a model was obtained by regressing the logarithm of the dataset cumulative death count as a function of time and using the median age, the logarithm of the data slope and the intercept of this regression in the potential energy function of each model. These probabilities could then be combined to compute the posterior expectation of all model parameters and their covariances. Explicitly, each term in the energy function is of the form
\begin{equation}
E(x_d|x_m,\sigma(x_m)) = \frac{1}{2} {(x_d-x_m)^2\over{\sigma(x_m)^2}} + \ln \sqrt{2\pi}\sigma(x_m)
\end{equation}
In this expression, $x_m$ is, in turn, equal to the mean of the logarithm of the slope of the regression of the log cumulative death count over  40 simulations of the model as a function of time after first death, the mean of the intercepts of the same regressions, and the ScaleMedAge ($\times 40$ y) parameter of the model. Similarly, $x_d$ is the corresponding quantity for the dataset, with $\sigma(x_m)$ the standard deviation of the model quantities. In the case of ScaleMedAge, $\sigma$ is the population weighted standard deviation of the median age over all countries. We assume uniform grids in units of days. In particular, the rate-like model parameter values are therefore uniform in $1/\mathrm{rate}$ units because the regression determining the slope is as a function of time measured in days. We checked that the choice of $14$ days after the first death did not affect our results. As alluded to in the previous subsection, the posterior expectation of the ScaleMedAge parameter ($\times40$ y) for each location should turn out to be close to the actual median age for each location in our setup, and this was achieved (right column, Figure \ref{fig:histogram}).

\subsection{Simulations}
We simulated our grid of models on the NIH Biowulf cluster. Our grid comprised of 56448 $\times 2$ parametrized models simulated with 40 random networks each and parameters in all possible combinations from the following lists:
%(units are per day for rates and years for all age measures, with scale parameters dimensionless): 
\begin{itemize}
\item ScalePopDens: $[2,2.5,3,3.5,4,4.5,5,5.5]$
\item InfRate: $[1/20,1/18,1/16,1/14,1/12,1/10,1/8]$ d${}^{-1}$
\item DeathRate: $[1/2000,1/1750,1/1500,1/1250,1/1000,1/750,1/500]$ d${}^{-1}$
\item RecRate: $ [1/31,1/28,1/25,1/22,1/19,1/16,1/13,1/10]$ d${}^{-1}$
\item AgeIncrease: $[5.4,5.5,5.6]$
\item ScaleMedAge: $[0.5,0.65,0.8, 0.95,1.1,1.25] \times 40$ y
%\item SusAge: $[0.] $y
\item $N: 9\cdot 10^4$ or $10^5$ nodes

\end{itemize}
These generated datasets had varying lengths depending on model parameters, with a total size of approximately 5Tb. It is important to note that at no point did we make use of the infection or recovery data, following the argumentation laid out in  \cite{lourenco2020fundamental}. We then matched the simulation results to three different datasets: US states, European Union countries, European Union countries along with Australia, Canada, Israel, Japan, South Korea, Taiwan, New Zealand, and the United Kingdom (denoted EU+ in the following) to determine posterior model probabilities given these datasets. 

%Finally, we simulated the most probable models from these analyses on even bigger networks ($5\cdot 10^5$ or $4\cdot 10^5$ nodes) to ensure that there is no size dependence in our results. 

\section{Results}
\label{results}

Figure \ref{fig:histogram} (left column) shows that only a small fraction of the 56448 $\times 2$ simulated parameter sets had a non-negligible posterior probability ($p>10^{-8}$): 1888 models for the US, 1367 models for the EU and 5969 models for the EU+ datasets.

%Figure \ref{fig:histogram} (left column) shows 
\begin{figure*}
\centering
\includegraphics[width=17cm]{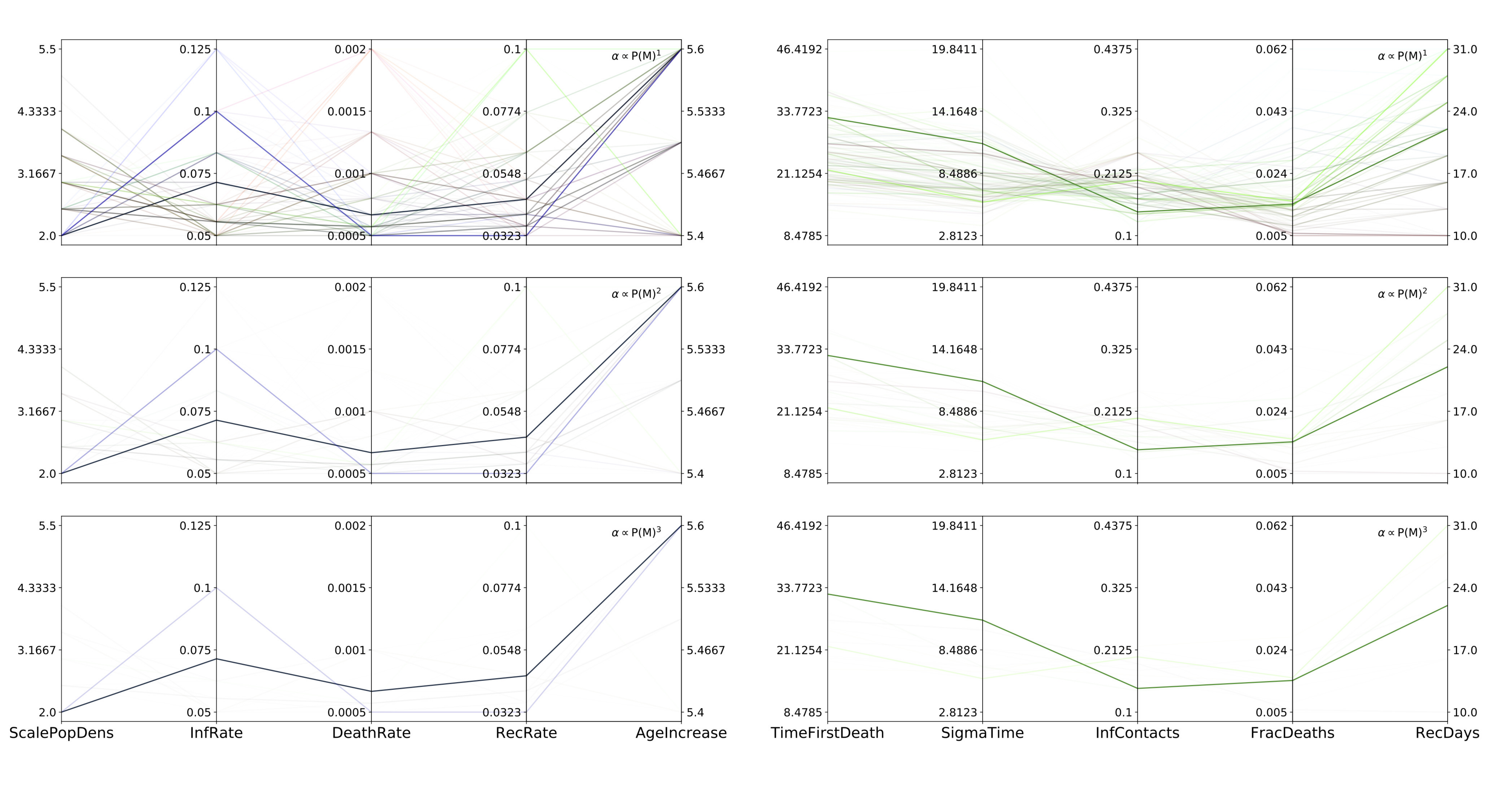}
\caption{ \label{fig:US}US states data: Parallel coordinate plots showing the parameters of models with varying posterior probabilities ($p>10^{-8}$) Left column: Model parameters. Each model corresponds to a single line across each plot. The RBG values of the lines are set by (normalized)(InfRate, DeathRate, RecRate) values of the corresponding model. For example, a model with high RecRate but low InfRate and low DeathRate is represented by a green line whereas a model with high InfRate and high DeathRate but low RecRate is represented by a purple line. Right column: Derived or simulated model values. The RBG values of the lines are set by (normalized)(InfContacts, FracDeaths, RecDays) values. In all plots, the opacity parameter $\alpha$ for each line is proportional to a power of the posterior probability of the corresponding model, $\alpha \propto P(M|D)^n, n=1,2,3.$ 
}
\end{figure*}
\begin{figure*}
\centering
\includegraphics[width=17cm]{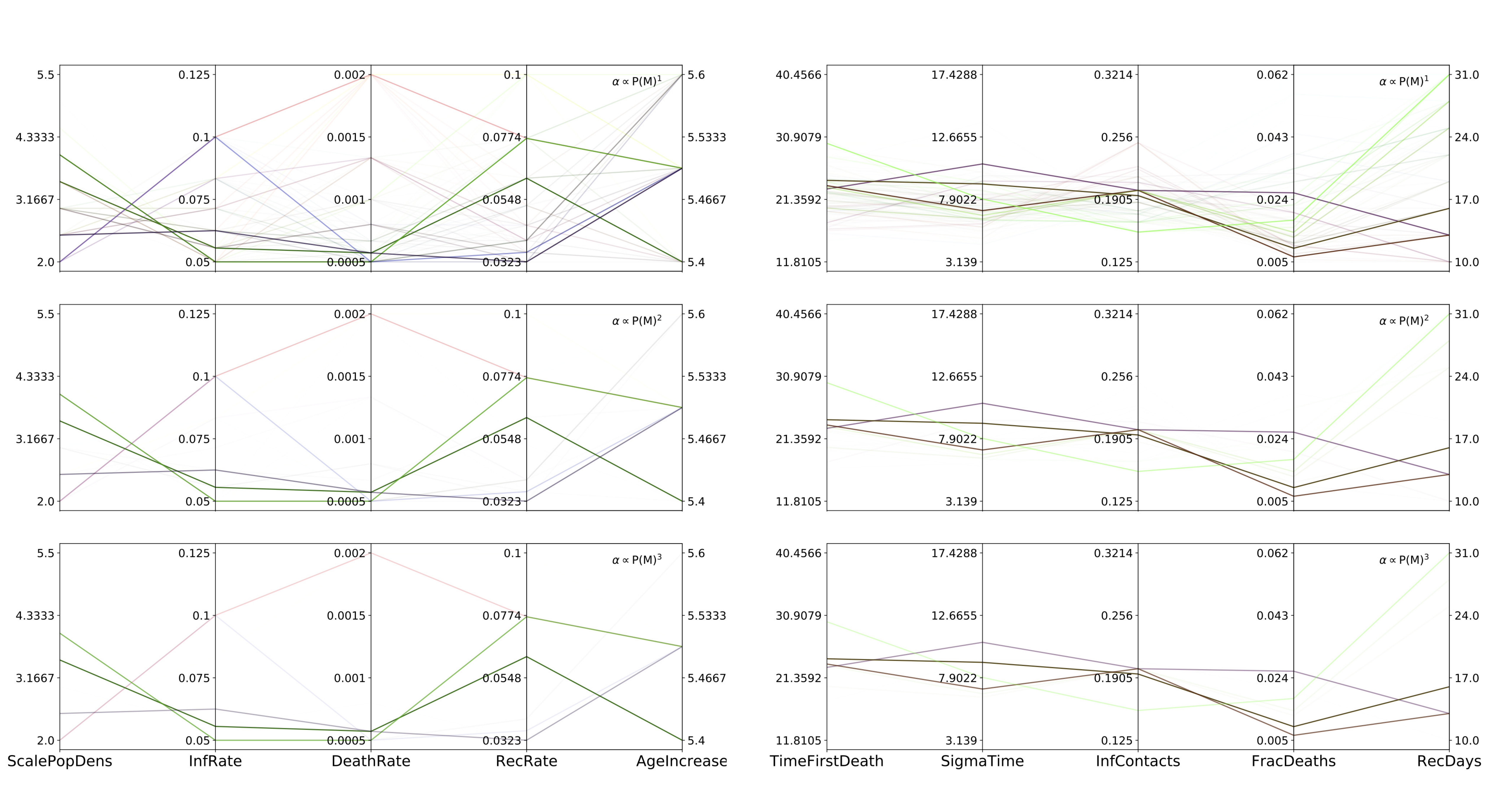}
\caption{ \label{fig:EU} European Union data: Parallel coordinate plots showing the parameters of models with varying posterior probabilities ($p>10^{-8}$) Left column: Model parameters.  Right column: Derived or simulated model values. (Colors and opacity as in Figure \ref{fig:US}.)}
\end{figure*}
\begin{figure*}
\centering 
\includegraphics[width=17cm]{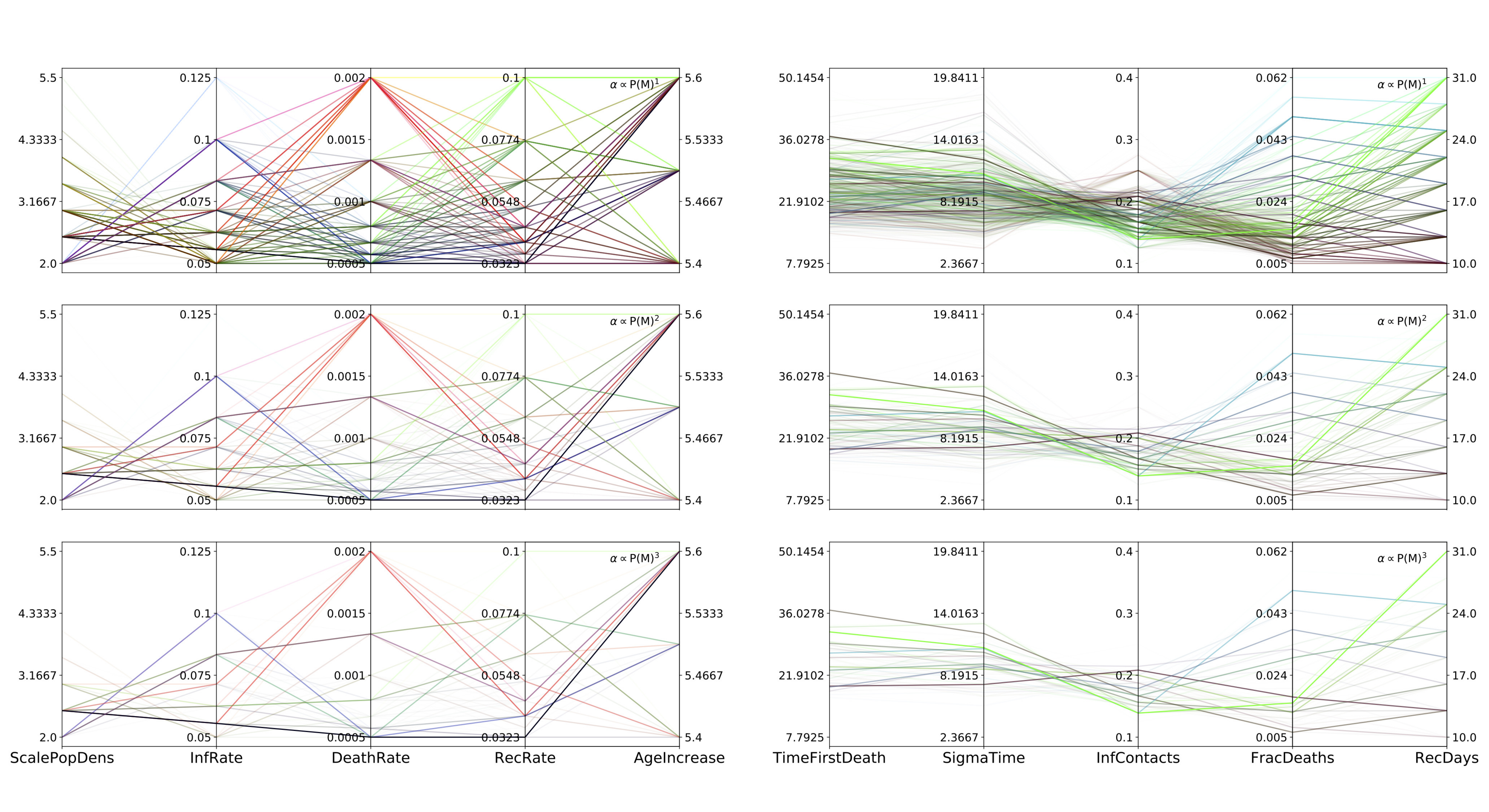}
\caption{ \label{fig:EUplus} European Union$+$ data: Parallel coordinate plots showing the parameters of models with varying posterior probabilities ($p>10^{-8}$) Left column: Model parameters. Right column: Derived or simulated model values. (Colors and opacity as in Figure \ref{fig:US}.)}
\end{figure*}
We use parallel coordinate plots to visualize the distributions of parameters in the models.  In these plots, each line corresponds to a model. The opacity of a line, $\alpha,$ is used to visualize the posterior probability of the model. By varying the dependence of this opacity $\alpha$ on increasing powers of the model posterior probability, we can gain a little intuition for the location of the mode(s) of the posterior distribution and the width of the distribution in the space of models. Going down each column, the dominant lines become those corresponding to the models with the highest posterior probability. The left columns of Figures \ref{fig:US} (US posterior), \ref{fig:EU} (EU posterior)  and \ref{fig:EUplus} (EU+ posterior) show parameters with colors weighted by the posterior probabilities of the models from our set of 56448 $\times 2$ models laid out on a five-dimensional grid and the right columns show derived or simulated quantities of interest, such as the time of first death, TimeFirstDeath, following the initial infection, the standard deviation of TimeFirstDeath within each stochastic model averaged over all posterior weighted models (SigmaTime), the network infectivity (InfContacts) defined as the product of the infection rate (InfRate) and the average number of network connections (ScalePopDens), the fraction of deaths per infection (FracDeaths), and the number of days an individual is contagious (RecDays). 

%Based on these results, we formulated a refined grid with some expanded limits (31500 models) to ensure that the results were not constrained overly by the models tested. This grid had the following values of parameters: 
%\begin{itemize}
%\item ScalePopDens: [2.0, 3.0, 4.0, 5.0, 6.0, 7.0]
%\item InfRate: [0.025, 0.035, 0.045, 0.055, 0.065]
%\item DeathRate: [0.005, 0.015, 0.025, 0.035, 0.045, 0.055]
%\item RecRate:  [0.005, 0.01, 0.015, 0.02, 0.025, 0.03, 0.035]
%\item AgeIncrease: [0.5, 1.0, 1.5, 2.0, 2.5]
%\item ScaleMedAge: [0.5, 0.75, 1.0, 1.25, 1.5] $\times 40$
%\end{itemize}
%
The posterior expectation values of parameters for these three sets of Bayesian analyses are shown in Table \ref{tb:post_par}.
\begin{table}[t]
\centering
\caption{\label{tb:post_par}Comparison of posterior expectation values of model parameters}
\begin{tabular}{|l|c|c|c|}
\hline
Parameter & US & EU & EU+ \\
\hline\hline
ScalePopDens & 2.70 $\pm$ 0.67 & 2.99$\pm$0.74 &2.68 $\pm$0.57 \\
\hline
InfRate (/d) & 0.069 $\pm$ 0.019 & 0.068$\pm$ 0.019& 0.068 $\pm$0.016  \\
\hline
DeathRate (/d)& 0.00078 $\pm$ 0.0004 &0.0009 $\pm$ 0.0006& 0.001 $\pm$ 0.0005 \\
\hline
RecRate (/d)& 0.050 $\pm$ 0.019 & 0.058 $\pm$ 0.021 & 0.056 $\pm$0.021 \\
\hline
AgeIncrease & 5.53 $\pm$ 0.08 & 5.49 $\pm$ 0.06 & 5.51 $\pm$ 0.08 \\
\hline
InfContacts & 0.18 $\pm$ 0.03 & 0.19 $\pm$ 0.02 & 0.18 $\pm$ 0.03 \\
\hline
FracDeaths & 0.017 $\pm$ 0.009 & 0.016 $\pm$ 0.009 & 0.02 $\pm$ 0.013 \\
\hline
RecDays (d) & 22 $\pm$ 6 & 20 $\pm$ 7 & 20 $\pm$ 7 \\
%\hline
%ScaleMedAge ($\times$40 y) & 1.0 $\pm$ 0.0 & 1.0 $\pm$ 0.0 & 1.0 $\pm$ 0.0 \\
%\hline
%SusAge (y)& 49 $\pm$ 8 & 50 (fixed) & 50 (fixed)\\
\hline
TimeFirstDeath TimeFirstDeath (d)& 25 ($\pm$ 6) $\pm$ 8 ($\pm$2)& 23 ($\pm$ 3)$\pm$ 8($\pm$1.5) & 8.07($\pm$ 2.2)$\pm$ 4.6 ($\pm$ 1.2)\\
\hline
Corr(InfRate,ScalePopDens) & -0.64 & -0.87 & -0.68 \\
\hline
\end{tabular}
\end{table}
In contrast to these posterior expectations, the parameter values  and model probabilities of the most likely models are shown in Table \ref{tb:post_ML}. These values are, of course, constrained by the coarse grid that we were able to simulate. 
\begin{table}[!h]
\centering
\caption{\label{tb:post_ML}Comparison of model parameters for the most likely models}
\begin{tabular}{|l|c|c|c|}
\hline
Parameter &US & EU & EU+ \\
\hline\hline
ScalePopDens & 2.0 & 3.5 & 2.5 \\
\hline
InfRate (/d) & 0.0714 & 0.056 & 0.056 \\
\hline
DeathRate (/d)& .00067  &0.0006 & 0.0005 \\
\hline
RecRate (/d)& 0.045 &  0.0625 & 0.032 \\
\hline
AgeIncrease & 5.6 & 5.5& 5.6  \\
\hline
InfContacts & 0.14& 0.19 & 0.139  \\
\hline
FracDeaths & 0.0147  & 0.009  & 0.016  \\
\hline
RecDays (d) & 22  & 16  & 31  \\
\hline
TimeFirstDeath (d)& 32.5 $\pm$ 11.2& 23 $\pm$ 10.6 & 31.8 $\pm$ 10.8\\
\hline
$-$Log(Prob(Model)) &2.09 & 1.93&3.77 \\
\hline
\end{tabular}
\end{table}
The models predict the number of links of each node to be about 2.7 with an uncertainty of about 0.7. This is consistent with the approximately 25-30\% uncertainties that are evident in the other parameters. In particular, note that the network infectivity (InfContacts) has a factor of two smaller uncertainty than either of its factors as these two parameters (InfRate and ScalePopDens) cooperate in the propagation of the contagion and therefore turn out to have a negative posterior weighted correlation coefficient (Table \ref{tb:post_par}). The concordance of posterior expectation values (Table \ref{tb:post_par}) between all three datasets is noteworthy, with the most likely parameter sets not showing the same level of agreement. The importance of Bayesian model averaging is therefore apparent. 

\begin{figure*}
\centering
\includegraphics[width=17cm]{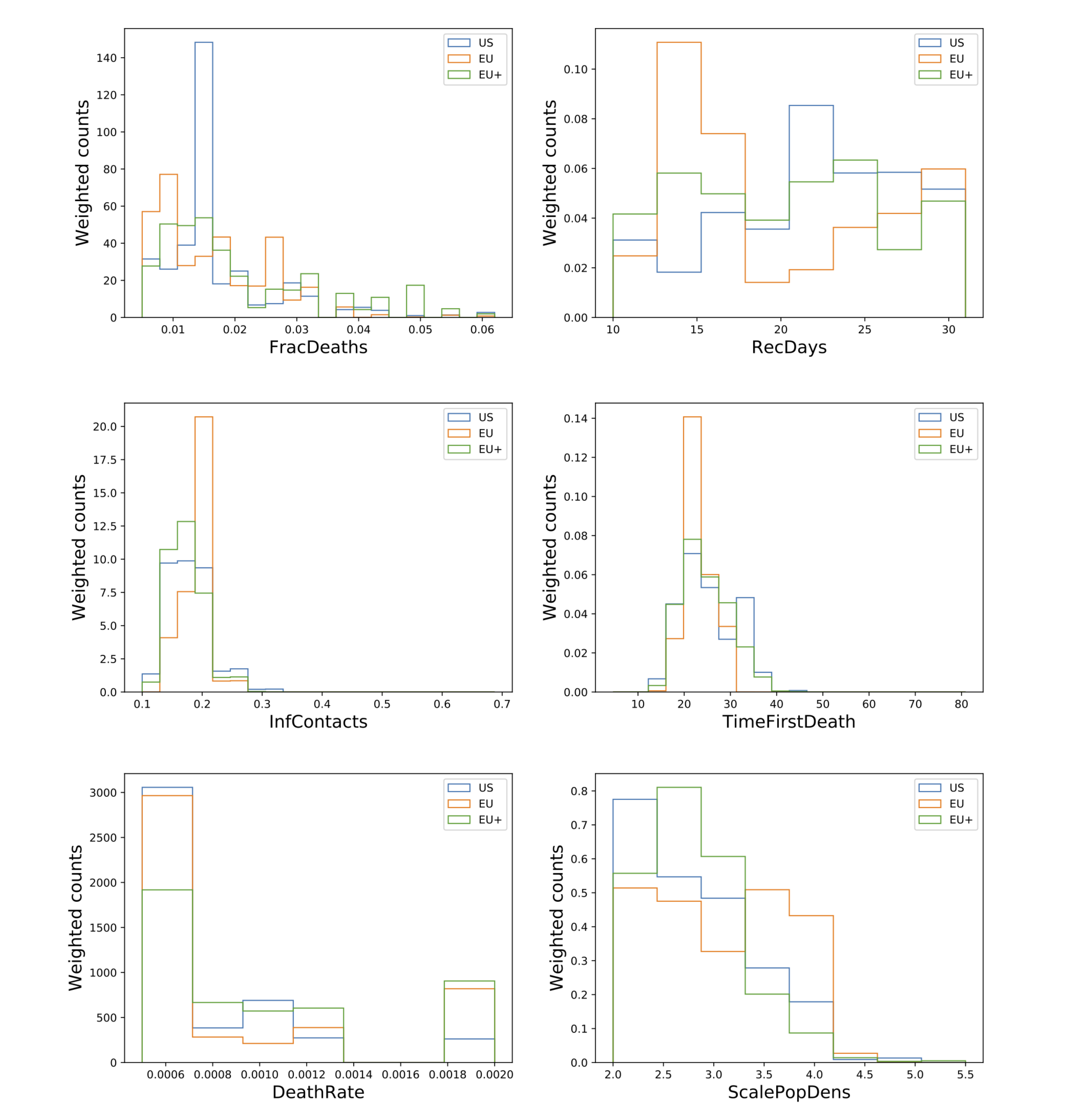}
\caption{ \label{fig:par_hist} Model posterior weighted distributions of the fraction of infected people that die (FracDeaths), number of days an individual is contagious (RecDays), network infectivity (InfContacts), time to first death following infection (TimeFirstDeath), death rate when infected (DeathRate) and the average number of edges connecting a node to other nodes (ScalePopDens) for all datasets (US, EU, EU+).
}
\end{figure*}

A more detailed view of the characteristics of the likely models in the form of  posterior probability weighted histograms of some derived parameters of interest is shown in Figure \ref{fig:par_hist}. The narrow widths of the distributions of FracDeaths, InfContacts and TimeFirstDeath make the larger spread of the RecDays contagious period particularly striking.

\begin{figure*}
\centering
\includegraphics[width=17cm]{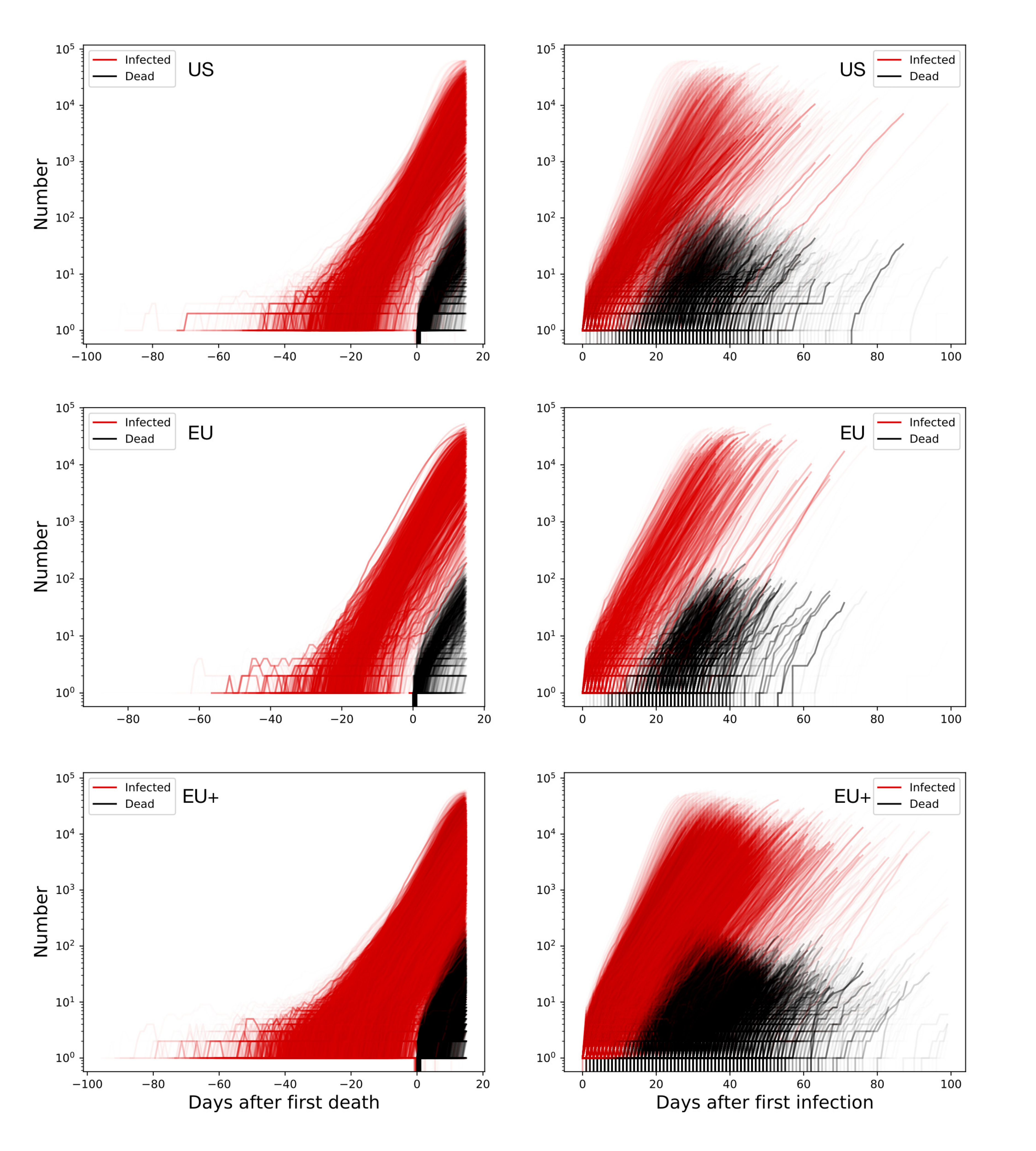}
\caption{ \label{fig:inf_dead} Model posterior weighted predictions of cumulative infections and deaths. Left column: Aligned with the time of first death. Right column: Aligned with the time of first infection. The opacity of a line $\alpha$ is weighted by the posterior probability of each model in that data set (US, EU, EU+).
}
\end{figure*}

The posterior weighted predictions of infected nodes and dead nodes are shown in Figure \ref{fig:inf_dead}. There is a remarkable range of more than two orders of magnitude in the number of infected nodes predicted 15 days after the first death but it should be noted that this spread is the consequence of exponential growth with the uncertainty $\pm0.03$ in InfContacts ($0.18$). This goes along with the approximately 80 day period between the first infection and the first death for a few outlier trajectories. However, it is also clear from the histograms in Figure \ref{fig:par_hist} and the mean TimeFirstDeath given in Table \ref{tb:post_par} that the likely value of this duration is considerably shorter.

\begin{figure*}
\centering
\includegraphics[width=15.5cm]{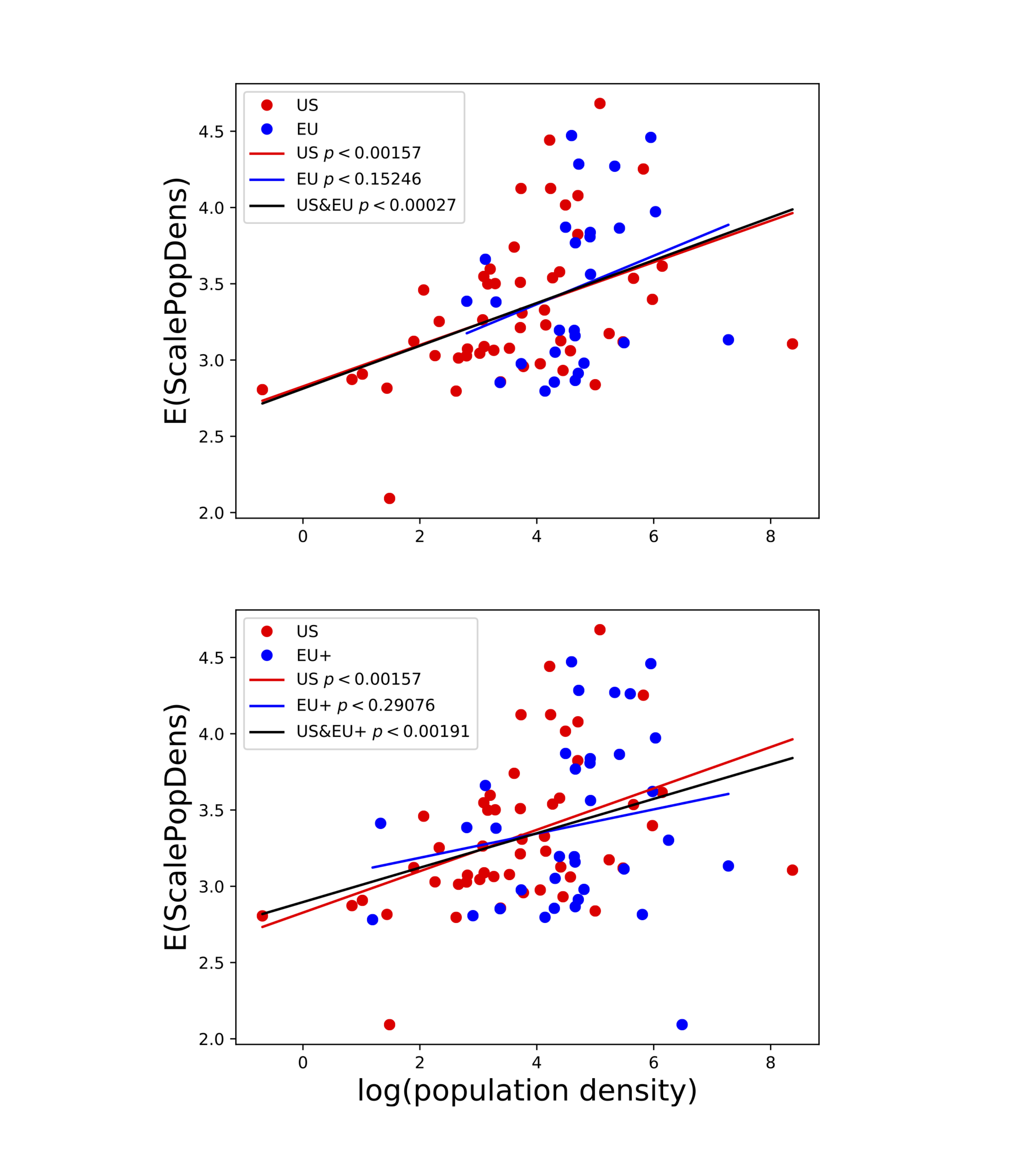}
\caption{ \label{fig:regress} Logarithm of population density (per $\mathrm{km}^2$) vs. the expectation value of the parameter ScalePopDens. }
\end{figure*}

Finally, we evaluated a possible correlation between the actual population density and the ScalePopDens parameter governing network connectivity. We found a significant correlation ($p<0.0016, r = 0.43$) between the logarithm of the population density and the posterior expectation of the ScalePopDens parameter, where the Bayes probability of any model is determined only with respect to data from that specific state in the US dataset (Figure~\ref{fig:regress}):
\begin{equation}
\mathrm{ScalePopDens} = 0.136 \ln(\mathrm{population\ per}\ \mathrm{km}^{2}) + 2.8 .
\label{eq:regress}
\end{equation}
When we added the European Union countries to this regression, we obtained ($p<0.0003, r = 0.40$) 
\begin{equation}
\mathrm{ScalePopDens(US \& EU)} = 0.140\ln(\mathrm{population\ per}\ \mathrm{km}^{2}) + 2.8 .
\label{eq:regress2}
\end{equation}
When we added additional countries to the European Union countries in this regression, we obtained ($p<0.0019, r = 0.33$) 
\begin{equation}
\mathrm{ScalePopDens(US \& EU+)} = 0.11\ln(\mathrm{population\ per}\ \mathrm{km}^{2}) + 2.9 .
\label{eq:regress3}
\end{equation}
Thus the network connectivity relevant to the spread of the pandemic is a function of the population density. 

\section{Discussion}
\label{discussion}

While epidemiology is not the standard stomping ground of statistical physics, Bayesian model comparison is naturally interpreted in a statistical physics context. We showed that taking this interpretation seriously leads to enormous reductions in computational effort. Given the complexity of translating the observed manifestations of the pandemic into an understanding of the virus's spread and the course of the infection, we opted for a simple data-driven approach, taking into account population age distributions and the age dependence of the death rate. While the conceptual basis of our approach is simple, there were computational difficulties we had to overcome to make the implementation amenable to computability with finite computational resources. Our results were checked to not depend on the size of the networks we simulated, on the number of stochastic runs we used for each model, nor on the number of days that we used for the linear regression. All the values we report in Table \ref{tb:post_par} are well within most estimated ranges in the literature \cite{berger2020seir, calvetti2020bayesian, Kucharski2020, zhang2020evolving, lourenco2020fundamental, chow2020global, chang2020modelling, keeling2020predictions, seydi2020unreported, raghavan2020using, hota2020closedloop, singh2020c19tranet, mao2020datadriven, liu2020distributed, susanto2020infect, whiteley2020inference, johansson2020masking, tuite2020mathematical, maltezos2020methodology, plata2020simulating, ganesan2020spatiotemporal, ohsawa2020stay, bastos2020covid19} but with the benefit of uncertainty estimates performed with a uniform model prior. While each location ball may range over a broad distribution of models, the consensus posterior distribution (Table \ref{tb:post_par}) shows remarkable concordance across datasets.
\begin{figure*}
\centering
\includegraphics[width=15.5cm]{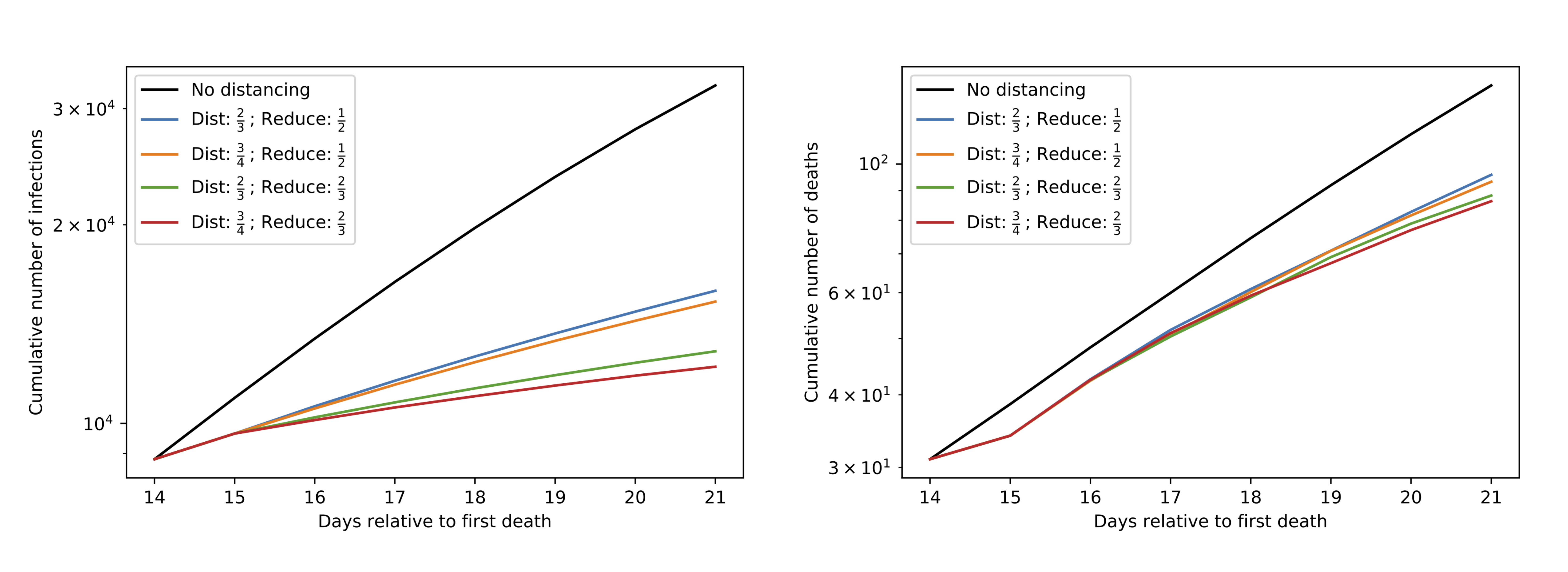}
\caption{ \label{fig:distancing} Lowering network connectivity has a larger effect on deaths than increasing compliance. Dist denotes the fraction of compliant nodes in the network and Reduce is the fraction of edges deleted at random from compliant nodes in the graph after social distancing is effectuated on Day 15 after the first death.}
\end{figure*}

We can predict the posterior distribution of time of initial infection, TimeFirstDeath, as shown in Table \ref{tb:post_par}. The dynamic model can predict the number of people infected after the first infection (right panel, Figure~\ref{fig:inf_dead}) and relative to the time of first death  (left panel, Figure~\ref{fig:inf_dead}) because we made no use of infection or recovery statistics in our analysis \cite{seydi2020unreported}. Note the enormous variation in the number of infections for the same parameter set, only partly due to stochasticity of the networks themselves, as can be seen by comparing the upper and lower rows of Figure~\ref{fig:simulations40}. 

With parameters intrinsic to the infection held fixed, we can predict the effect of various degrees of social distancing by varying network connectivity. We assumed that a certain fraction of nodes in the network would comply with social distancing and only these compliant nodes would reduce their connections at random by a certain fraction. Figure~\ref{fig:distancing} shows the effects of four such combinations of compliant node fraction and fraction of contact reduction by the compliant nodes. Interestingly, while the fraction of compliant nodes that implement social distancing is relevant to the flattening of the cumulative number of deaths, the more important parameter appears to be the degree of social distancing by the compliant nodes, which in the present context means the fraction of edges in the network that are randomly deleted for compliant nodes only to simulate social distancing. The rate of exponential increase in the number of infected individuals is $0.18\pm .03$ per day without social distancing. This rate falls to about $0.08$ per day
with even the most pessimistic social-distancing effectuation shown in Figure~\ref{fig:distancing}. While the uncertainty of $0.03$ may not appear to be large, it appears in an exponent and leads to enormous ranges in the predicted number of infected individuals, a subject of great interest given possible asymptomatic transmission of the virus\cite{seydi2020unreported}. With this caveat in mind, just focusing on the mean value $0.18$ per day implies that such a contact reduction corresponds to 65 million vs. 166000  infected individuals after a period of 100 days starting from one infected individual if social distancing is initiated two weeks after the first death on day 25.

As is apparent in contrasting Figures~\ref{fig:US}, \ref{fig:EU} and \ref{fig:EUplus}, the mode of the posterior distribution is not particularly representative of the distribution of parameters. Comparing the parameter values of the most likely model (Table \ref{tb:post_ML}) with the posterior expectations of parameters (Table \ref{tb:post_par}) shows that the Bayes entropy of the model posterior distribution is an important factor to consider, validating our initial intuition that optimization of model parameters would be inappropriate in this analysis. The regression we found (Eq.'s~\ref{eq:regress}, \ref{eq:regress2}, \ref{eq:regress3}) with respect to population density must be considered in light of the fact that many outbreaks are occurring in urban areas so they are not necessarily reflective of the true population density dependence. Furthermore, we did not find a significant regression for the countries of the European Union by themselves, perhaps because they have a smaller range of population densities, though the addition of these countries into the US states data further reduced the regression $p$-value of the null hypothesis without materially altering regression parameters. Detailed epidemiological data could be used to clarify its significance. 

\cite{reich2020modeling, endo2020estimating, furuse2020clusters, lau2020characterizing} have suggested the importance of super-spreader events but we did not encounter any difficulty in modeling the available data with garden variety $G(n,p)$ networks. Certainly if the network has clusters of older nodes, there will be abrupt jumps in the cumulative death count as the infection spreads through the network. Furthermore, it would be interesting to consider how to make the basic model useful for more heterogenous datasets such as all countries of the world with vastly different reporting of death statistics. Using the posterior distribution we derived as a starting point for more complicated models may be an approach worth investigating.

Infectious disease modeling is a deep field with many sophisticated approaches in use \cite{namatame2016agent,liu2017infectious,kiss2017mathematics,y2018charting} and, clearly, our analysis is only scratching the surface of the problem at hand. Network structure, in particular, is a topic that has received much attention in social network research \cite{Anderson1999,Welch2011,Orsini2015,amblard2015models,becker2015modeling}. Bayesian approaches have been used in epidemics on networks modeling \cite{groendyke2011bayesian} and have also been used in the present pandemic context in  \cite{reich2020modeling,calvetti2020bayesian,karaivanov2020social}.  To our knowledge, there is no work in the published literature that has taken the approach adopted in this paper. %Our contribution should be taken as a small piece in the larger puzzle of pandemic containment strategies.

There are many caveats to any modeling attempt with data this heterogenous and complex. First of all, any model is only as good as the data incorporated and unreported SARS-CoV-2 deaths would  impact the validity of our results. Secondly, if the initial deaths occur in specific locations such as old-age care centers, our modeling will over-estimate the death rate. A safeguard against this is that the diversity of locations we used may compensate to a limited extent. Detailed analysis of network structure from contact tracing can be used to correct for this if such data is available, and our posterior model probabilities could guide such refinement. Thirdly, while we ensured that our results did not depend on our model ranges as far as practicable, we cannot guarantee that a model with parameters outside our ranges could not be a more accurate model. The transparency of our analysis and the simplicity of our assumptions may be helpful in this regard. All code is available \cite{code}.

\section*{Acknowledgment}
We are grateful to Arthur Sherman for helpful comments and questions and to Carson Chow for prepublication access to his group's work \cite{chow2020global}. This work was supported by the Intramural Research Program of the National Institute of Diabetes and Digestive and Kidney Diseases, NIH. This work utilized the computational resources of the NIH HPC Biowulf cluster. (http://hpc.nih.gov)
\bibliography{covid}
\end{document}